\documentclass[iop]{emulateapj}
\usepackage{apjfonts}

\newcommand{\galex}{{\it GALEX}}  
\newcommand{\hst}{{\it HST}}  
\newcommand{\ha}{{H$\alpha$}}

\journalinfo{ApJ Letters, accepted}

\shorttitle{STAR FORMATION IN EARLY-TYPE GALAXIES}
\shortauthors{SALIM \& RICH}

\begin{document}

\title{Star Formation Signatures in Optically Quiescent 
Early-type Galaxies} 

\author{Samir Salim} 
\affil{Department of Astronomy, Indiana University,
  Bloomington, IN 47404, salims@indiana.edu} 
\and \author{R.\ Michael Rich} 
\affil{Department of Physics and Astronomy, University of
  California, Los Angeles, CA 90095}

\begin{abstract}
In recent years argument has been made that a high fraction of
early-type galaxies in the local universe experience low levels
($\lesssim 1 M_{\odot} {\rm yr}^{-1}$) of star formation (SF) that
causes strong excess in UV flux, yet leaves the optical colors
red. Many of these studies were based on \galex\ imaging of SDSS
galaxies ($z \sim 0.1$), and were thus limited by its $5''$ FWHM. Poor
UV resolution left other possibilities for UV excess open, such as the
old populations or an AGN. Here we study high-resolution
far-ultraviolet \hst/ACS images of optically quiescent early-type
galaxies with strong UV excess. The new images show that
three-quarters of these moderately massive ($\sim 5\times
10^{10}M_{\odot}$) early-type galaxies shows clear evidence of
extended SF, usually in form of wide or concentric UV rings, and in
some cases, striking spiral arms. SDSS spectra probably miss these
features due to small fiber size. UV-excess early-type galaxies have
on average less dust and larger UV sizes ($D>40$ kpc) than other
green-valley galaxies, which argues for an external origin for the gas
that is driving the SF. Thus, most of these galaxies appear
`rejuvenated' (e.g., through minor gas-rich mergers or IGM accretion).
For a smaller subset of the sample, the declining SF (from the
original internal gas) cannot be ruled out. SF is rare in very massive
early-types ($M_*>10^{11} M_{\odot}$), a possible consequence of AGN
feedback. In addition to extended UV emission, many galaxies show a
compact central source, which may be a weak, optically inconspicuous
AGN.

\end{abstract}

\keywords{galaxies: evolution---ultraviolet: galaxies---galaxies:
  elliptical and lenticular, cD}

\section{Introduction}

In this paper we present a discovery of significantly extended regions
of star formation in some early-type galaxies (ETGs)--galaxies usually
thought to lie on the passive side of galaxy bimodality. Bimodality in
terms of morphology and color has been known since the earliest
studies of galaxies, but it was not until the massive datasets of the
Sloan Digital Sky Survey (SDSS) that fuller implications in terms of
galaxy evolution became evident \citep{strateva,kauff03}. Optical
colors reflect the mean age of stellar populations and are therefore
sensitive only to high levels of continuous SF \citep{kauff07}. If the
SFR per unit stellar mass (specific SFR) drops below some threshold,
optical colors become uniformly red and SDSS photometry cannot
distinguish a truly passive galaxy from one that also contains a young
population. These limitations are alleviated when ultraviolet (UV)
photometry, dominated by young stars, is available. Early results from
\galex\ showed that a surprisingly high fraction (15\%) of optically
red SDSS ETGs exhibit strong UV excess \citep{yi}. \citet{rich} found
strong far-UV (FUV) excess even when selecting ETGs with no
\ha\ emission in SDSS spectra. Is this UV excess due to star
formation, as assumed by \citet{yi,kaviraj07,kaviraj09}? While SF and
molecular gas have been studied in nearby early-type galaxies for some
time (e.g., \citealt{bregman}), their significance as a possible {\it
  phase} in galaxy evolution or a {\it mode} of galaxy growth requires
the large samples we have today. Before considering such far-reaching
implications one must ask whether other explanations for the UV flux
exist? After all, nearby ellipticals are known to exhibit a moderate
UV excess (the ``UV upturn'', \citealt{code}), that comes from old
stars (presumably hot horizontal branch), and not massive young stars
\citep{greggio}. Also, a continuum from a weak AGN could in principle
produce an UV excess \citep{agueros}.

With $5''$ FWHM, \galex\ imaging marginally resolves SDSS galaxies at
$z \sim 0.1$ (angular diameter $<20''$), which is why we turned to
{\it high-resolution} FUV imaging with the Solar Blind Channel (SBC)
of the ACS. \hst\ images of our sample of massive ETGs with strong UV
excess and no obvious optical signs of SF reveal a surprise:
they are dominated by {\it extended} star formation on scales of
10--50 kpc, and with rates of up to 2 $M_{\odot} {\rm yr^{-1}}$.

\section{Data and Sample}

Our sample is selected from the SDSS DR4 main spectroscopic survey
($r\leq 17.77$) matched to \galex\ Medium Imaging Survey IR1.1 (${\rm
  FUV_{lim}}=22.7$; AB magnitudes throughout). The details of SDSS and
\galex\ data and the matching procedure are given in \citet{s07}.

\begin{figure}
\epsscale{1.2} \plotone{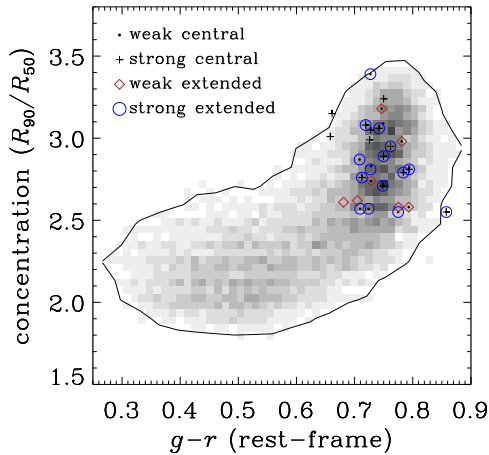}
\caption{Optical light concentration against optical color. In all
  figures the strong UV-excess sample is shown with symbols; underlying
  galaxies as greyscale, outlined by 90\% contour. The sample is selected to be high
  concentration, but no explicit optical color selection was applied.}
\label{fig:conc}
\end{figure}

From the matched SDSS/\galex\ catalog we select optically quiescent
early-type galaxies (QETGs) in the following way: (1) redshift $<0.12$
to yield a sample with larger angular sizes, (2) $i$-band light
concentration (ratio of 90 and 50\% Petrosian radii) $>2.5$ to select
dominant spheroids (Fig.\ \ref{fig:conc}), (3) no detectable
\ha\ emission based on DR4 version of \citealt{b04} BPT classification
(``No \ha'' category in \citealt{s07}). Note that no color selection
has been applied to select QETGs.

\begin{figure}
\epsscale{1.2} \plotone{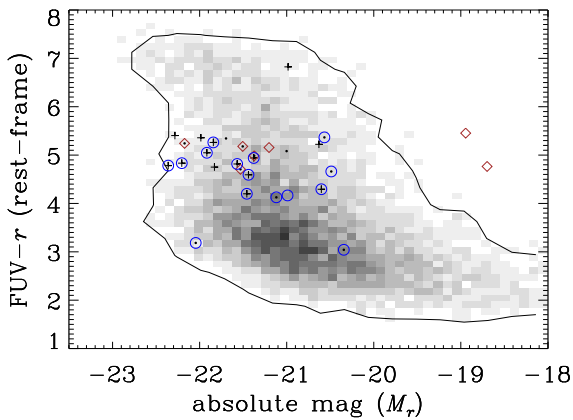}
\caption{UV-optical color-magnitude diagram in ${\rm FUV}-r$. The
  sample is selected to be bluer than typical ETGs, yet optical colors
  are red (Fig.\ \ref{fig:conc}).}
\label{fig:cmd}
\end{figure}

Out of $\sim 1000$ QETGs, one-fifth has rest-frame FUV-optical colors
(Fig.\ \ref{fig:cmd}) bluer than ${\rm FUV}-r = 6.0$\footnote{Colors
  are given in rest frame throughout.}. \citet{donas} show that nearby
RC3 ellipticals (without lenticulars), where the UV excess is known to
come from classical UV upturn (old populations), are redder than this
limit. In contrast, we are interested in QETGs with {\it strong} UV
excess, so we select galaxies with ${\rm FUV}-r < 5.3$. There are 60
such galaxies from which we exclude blends, obvious optical
disturbances, late-type contaminants, and E+A post-starbursts (based
on H$\delta_A$ index), to arrive at a final \hst\ sample of 30.

The UV-optical (${\rm FUV}-r$) color range of our sample can be seen
from Figure \ref{fig:cmd}. The sample is presented with symbols, while
greyscale represents all SDSS-\galex\ galaxies at $z<0.12$ (the
underlying population). By selection, the sample is bluer than the
${\rm FUV}-r$ red sequence, with two galaxies ( ${\rm FUV}-r \approx
3$) lying squarely in the ${\rm FUV}-r$ blue sequence.\footnote{When
  referring to blue/red sequence/cloud or the green valley, one needs
  to specify the color used to distinguish these populations. E.g.,
  the green valley makes sense as a separate population only in
  UV-optical colors \citep{wyder}}.

While no optical color cut has been explicitly applied, our strong UV
excess sample has distinctly {\it red} optical color
(Fig.\ \ref{fig:conc}), placing them firmly in the optical red
sequence. Similar confinement to the red sequence is seen in $u-r$
colors (not shown). \citet{sch09} and \citep{kannappan} study blue
early-type galaxies in SDSS selected by atypically blue
$u-r$. However, our sample galaxies are {\it redder} than their $u-r$
cuts.

\section{\hst\ UV observations}

With FWHM of $5''$, \galex\ makes it difficult to pinpoint the origin
of the FUV light at $z\sim0.1$. In contrast, ACS/SBC on the
\hst\ places 80\% of point source energy in $0\farcs4$
\citep{acsford}. Targets were observed with one orbit through the
long-pass filter F125LP ($\lambda_{\rm eff}=1459$ \AA). Processing was
performed using MULTIDRIZZLE with SBC-optimized parameters, and
smoothed using ADAPTSMOOTH \citep{adapt}.

\section{Results}

\subsection{UV morphology}

Twenty-nine targets were successfully imaged, and each produced a
detection, either of a compact central source or of extended
structures (or both). To our surprise, 22 galaxies (76\%) revealed an
extended UV morphology and an additional three had UV patches within
several arcsec of the nucleus. In all cases the extended UV emission
is structured, and thus results from {\it star formation} (unlike a
diffuse component one might expect from an old population). The UV
extent is typically larger than the optical size, though mostly
contained within a radius containing 90\% {\it optical} Petrosian
flux. We divide extended structures into strong (15) and weak (7)
based on visual appearance and provisional flux measurements. These
are labeled in figures by circles and diamonds respectively. In all
galaxies save four a compact central source is present as well, which
we divide into strong (14; plus sign) and weak point sources (11;
dots).

\begin{figure*}
\epsscale{1.0} \plotone{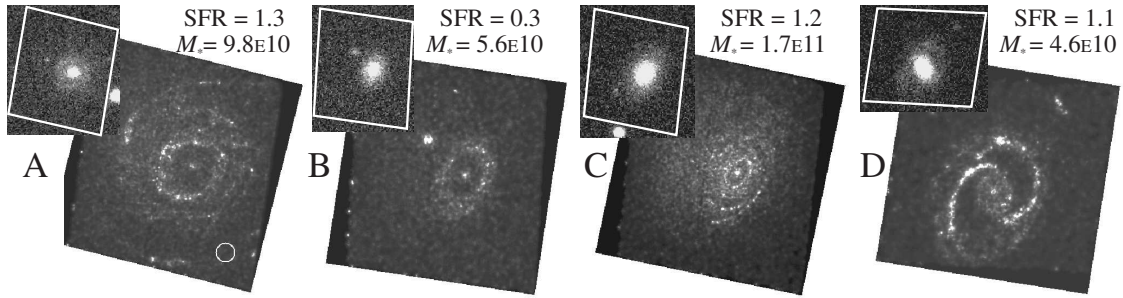}
\caption{\hst\ ACS/SBC far-UV images of several strong UV-excess
  galaxies. Insets show SDSS $g$-band image. ACS images are $\sim
  33''$ across ($\sim 63$ kpc). SFRs are given in $M_{\odot}{\rm
    yr}^{-1}$ and stellar masses in $M_{\odot}$.}
\label{fig:images}
\end{figure*}

Several strong extended UV morphologies are illustrated in Figure
\ref{fig:images}. Galaxy (a) has a weak central source, while others
exhibit a strong central source. Insets show $g$-band SDSS images with
ACS fields indicated. As required by sample selection, optical images
appear like ETGs, though, after the fact, hints of structure
are visible in (a), (d) and a couple of other galaxies.

The galaxy in panel (a) of Figure \ref{fig:images} shows multiple star
forming features, including an inner ring and what look like
flocculent spiral arms. Two other galaxies in the sample show
flocculent features. Panel (b) shows a galaxy with a wide SF
ring. Such morphology is most common in our sample (eight other
cases). Next is one of four galaxies showing thin concentric
rings. The last (panel d) is the most striking in terms of surface
brightness, with two bright spiral arms emanating from what appears to
be a co-rotation ring, connected by a fainter outer Lindblad resonance
ring \citep{schwarz}. The optical image shows a hint of red spiral
arms. Two other galaxies in the sample appear to have spiral arms,
albeit less prominent.

\subsection{Placing the UV excess sample in context}

We first return to the plot showing concentration vs.\ optical color
(Fig.\ \ref{fig:conc}). Two of the optically bluest sources do not
have extended structures, emphasizing the disconnect between the UV
excess and the optical colors. On average, galaxies with a strong
central source tend to be more optically concentrated than those with
only a weak central source, which are in turn more highly concentrated
than those with no central source.

Figure \ref{fig:cmd} shows the ${\rm FUV}-r$ color-magnitude
diagram. Sample galaxies span a wide range in absolute magnitude, with
a median of $M_r = -21.4$, which is close to $M_r^*=-21.6$
\citep{blanton}. Similarly to concentration, there exists the trend
between the strength of central source and absolute magnitude. Not
surprisingly, this trend is also present vs.\ stellar mass and, even
more strongly, velocity dispersion. Compared to all UV-detected QETGs,
the UV-strong ones have somewhat lower stellar masses (median is $\log
M_*=10.7$) , with a clear deficit at $\log M_* > 11$. In other words,
most massive ETGs do not exhibit strong UV excess. \footnote{Chabrier
  IMF stellar masses, star formation rates (SFRs) and FUV attenuations
  were derived from UV/optical SED fitting described in \citet{s07}.}
Two galaxies stand out as low luminosity (and mass; $\log M_* \approx
9.6$). They exhibit less extended (yet still patchy) UV morphology
without a central source.

\begin{figure}
\epsscale{1.2} \plotone{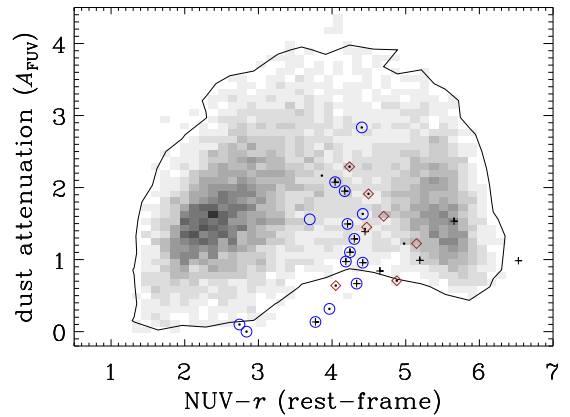}
\caption{FUV dust attenuation vs.\ UV-optical color. Sample galaxies
  fall in the green valley, but have lower dust attenuations than
  other such galaxies.}
\label{fig:afuv}
\end{figure}

Switching from ${\rm FUV}-r$ to the less dust-sensitive ${\rm NUV}-r$
will allow us to place the sample in a better SF history context.
Figure \ref{fig:afuv} plots FUV attenuation against ${\rm NUV}-r$
color. First focusing on the color, we see that the majority of UV
excess ETGs fall in the green valley ($4<{\rm NUV}-r<4.5$).  ${\rm
  NUV}-r$ basically ``resolves'' the {\it optical} red sequence into a
green valley and {\it UV-optical} red sequence. Every red galaxy had
to pass through the green valley at some point, and presumably most
green valley galaxies are transiting for the first time
\citep{martin}. After a galaxy gets on a red sequence it can enter the
green valley again if it experiences a star formation episode
(``rejuvenation''). From Figure \ref{fig:afuv} one notices that sample
galaxies in the UV red sequence lack strong extended structures, while
the contrary is true for bluer galaxies. Thus, we conclude that UV
excess galaxies are powered by star formation if they are in the green
valley or bluer, but they are dominated by a central source (perhaps
an AGN) if despite blue ${\rm FUV}-r$ their ${\rm NUV}-r$ places them
in the red sequence. The dust attenuation of sample galaxies is smaller
than that of other galaxies of the same color, which may hold clues to
their evolutionary path. Different UV morphologies have no significant
differences in terms of dust attenuation.

\begin{figure}
\epsscale{1.2} \plotone{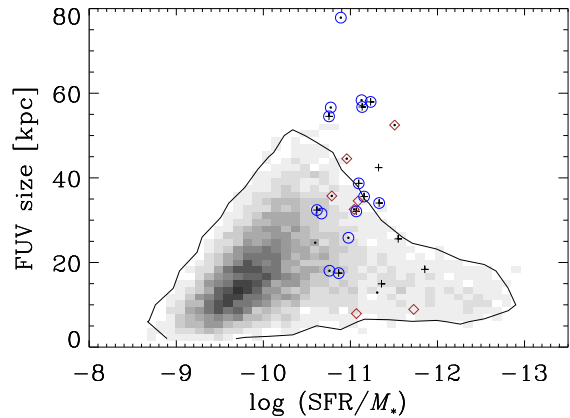}
\caption{FWHM far-UV physical size (from \galex) vs.\ specific SFR. UV
  excess sample galaxies are on average larger than other transitional
  galaxies.}
\label{fig:fwhm}
\end{figure}

To distinguish the effects of dust from those of SF history, we switch
from ${\rm NUV}-r$ to (dust-corrected) specific SFR (${\rm
  SFR}/M_*$). The specific SFR is plotted against the far-UV physical size
in Figure \ref{fig:fwhm}, with more actively star-forming galaxies on
the left. Our sample now spans a narrower range in ${\rm SFR}/M_*$
than it did in ${\rm NUV}-r$. Two galaxies that sat in the blue
sequence owed such blue colors to very low dust attenuation, and
actually have ``transitional'' specific SFRs as do other sample
galaxies. Galaxies without extended UV emission tend to have lower
specific SFRs, in agreement with inferences from color. Sample
galaxies span a range of SFRs from 0.01 to 2 $M_{\odot} {\rm
  yr^{-1}}$, with a median of 0.4 $M_{\odot} {\rm yr^{-1}}$.

UV sizes in Figure \ref{fig:fwhm} are \galex\ FWHM measurements for
both the sample and the underlying population (to avoid biasing one
measurement). Bursty galaxies are typically less massive and smaller
compared to those with $\log {\rm SFR}/M_*)=-10$. Beyond that, SF
activity decreases as do the UV sizes. However, sample galaxies are on
average significantly larger than the underlying galaxy sample of the
same specific SFR.

\section{Discussion}

This study was motivated by the results of \citet{rich} in which QETGs
were selected to study the classical, old-population UV upturn, but
instead resulted in much bluer samples that did not adhere to
relations expected for classical UV upturn. \citet{agueros} suggested
that the blue UV-optical extent came from a low-level AGN
contamination, while other studies (e.g., \citealt{yi}) attributed
blue colors in ETGs to star formation.

Our current study clearly demonstrates that the majority of UV excess
ETGs (even though optically quiescent) indeed have UV morphology
consistent with SF. This is especially the case for galaxies with
${\rm NUV}-r<5$, a cut originally proposed by \citet{yi} to
distinguish SF from the old-population UV upturn. So why is there
little or no \ha\, when from the FUV flux one expects \ha\ S/N of
around 40?  As Figure \ref{fig:images} shows, the $3''$ SDSS fiber may
entirely miss the UV-bright regions. However, in some cases there
appears to be UV emission within the fiber diameter and still no
\ha. Thus, there is a possibility that \ha\ emission is genuinely
suppressed in the central parts or throughout these galaxies. This
would happen if no SF took place over the last 10 Myr (the lifetime of
massive stars producing ionizing emission), or if few such stars are
being produced (e.g., \citealt{jlee}). Planned resolved \ha\ imaging
should settle this issue.

Now that we established the nature of the UV emission, we turn to the
origin of star-forming gas. We consider these scenarios: (1) observed
SF is what remains of the original, continuous SF; (2) SF results from
a gas accreted from a minor gas-rich merger or (3) accreted directly
from IGM, after a period of quiescence. In scenarios 2 and 3 galaxy is
again in the green valley after it was already on the red sequence,
while in 1 it is transiting from blue to red sequence for the first
time. While we selected the sample to have highly concentrated light
profiles, high concentrations are typical for most green valley
galaxies, so this cannot be used to distinguish scenario 1 from the
other two. The fact that strong UV-excess ETGs typically have blue
sequence-like masses ($\log M_* < 11$) may suggest scenario 1, because
if already passive ETGs are being rejuvenated, why would this happen
less often to massive ETGs?  SAURON sees a similar phenomenon in
nearby ETGs, where slow rotators do not show infrared signatures of SF
\citep{shapiro}. Presence/absence of SF may be another form of 'E-E'
dichotomy \citep{kormendy}, in which more luminous ($M_V <-21.5\pm 1$)
ellipticals are slow rotators and have X-ray emitting gas, possibly
maintained by radio mode AGN \citep{croton}, which prevents gas
accretion. In our sample, ETGs with SF all have $M_V > -22.0$.

Under scenario 1 our UV-strong galaxies will have properties like
other green-valley galaxies, most of which presumably transition from
blue to red and are not rejuvenated red galaxies
\citep{martin}. However, both low dust attenuation
(Fig.\ \ref{fig:afuv}) and large UV size (Fig.\ \ref{fig:fwhm}) argue
that this cannot be the case for the majority of the sample. Low dust
attenuation makes our sample more similar to ${\rm NUV}-r$ red
sequence than to green-valley galaxies. And if red galaxies acquired
gas from dwarfs, merging would not increase the dust content
significantly, since dwarfs are almost dust-free (e.g.,
\citealt{dale}). Similarly, the UV sizes of our sample are larger than
of the underlying galaxies with similar specific SFR, suggesting an
external origin. Together, this implies that most (but possibly not
all) of UV-excess ETGs are rejuvenated systems.

Rings, in one form or other, are present in 15 out of 22 galaxies
having extended UV. Rings in disk galaxies are commonly attributed to
either bar-induced resonances \citep{butacombes} or head-on collisions
\citep{appleton}. However, collisional rings, in addition to being
rare, are usually off-center \citep{lynds}, which is not the case
here. In a nearby SAURON elliptical NGC2974, UV rings detected by
\galex\ were explained as resonant features due to a hypothetical,
optically undetected bar \citep{jeong}. If ring structures in our
sample are due to resonances, the bars that induce them would need to
be more common than observed in S0 galaxies ($30\%$,
\citealt{aguerri}). Of galaxies in the sample, the galaxy in Figure
\ref{fig:images} (d) most clearly shows classical resonance features,
and the bar is present. Other ringed galaxies show no obvious evidence
of bars in SDSS images.  Perhaps some other, less-investigated process
introduces rotational asymmetry and forms rings
\citep{butacombes}. Resonances have been classically studied in disk
galaxies with existing gas, while here we are probably dealing with
newly-accreting gas, which may behave differently. Regardless of the
origin of the rings, we still need to identify the source of the cold
gas.

Recently, \citet{kaviraj09} advocated that minor gas-rich mergers
(scenario 2) can fully explain the extent and frequency of NUV-excess
ETGs. They also construct N-body simulations \citep{peirani} of minor
mergers, but these do not produce ring-like UV morphologies that we
see. Instead, their young stars are highly concentrated in the nuclear
region of the host. However, their hosts are kinematically more
similar to ellipticals, while as found by \citet{shapiro} and
indicated here, the only type of ETGs that allow extensive SF may be
S0/fast-rotators. Also, our selection against \ha\ in the central
region may bias against central starbursts. \citet{peirani}
simulations also predict that there should exist {\it stellar}
signatures of mergers, such as shells and tidal tails, but the current
optical data are not adequate to search for them.

The SFRs of our ETGs are similar to those of gas-rich dwarfs with
stellar 20 times lower masses. Ignoring dwarf galaxy duty cycles and
merger timescales, this would imply very minor mergers.

Finally, regarding scenario 3 (IGM accretion, \citealt{keres}), we
first ask why does IGM lead to SF rejuvenation and not just
the continuation of the original SF that formed the bulk of stellar
population. Many recent models require AGN feedback both to shut down
SF and to maintain their quiescence \citep{croton}. However, if the
AGN stopped being active, a galaxy could possibly re-acquire cold gas,
producing cyclical phases of AGN feedback and star formation. Our
galaxies, showing at best only very weak AGN activity, may currently
be in the renewed star-forming phase. (In contrast, in more massive
ETGs, $M_*>10^{11} M_{\odot}$, the accretion may be permanently
disrupted.) Large UV sizes (Fig.\ \ref{fig:fwhm}) and similarities
with XUV-disk phenomenon \citep{thilker,donovan} point towards this
scenario, but we currently cannot strongly favor IGM accretion over
minor mergers as the main source of gas in our sample.

In closing we turn to the nature of the UV emission of the compact
{\it central} source. While the compactness of UV emission suggests an
AGN, ACS/SBC resolution cannot outright rule out a highly nucleated
stellar bulge. We see a correlation between the strength of the
central source and velocity dispersion that may (through accretion
rate being a function of $\sigma$; \citealt{martin}) suggest an AGN,
but again does not exclude stellar origin. While emission lines are
too weak (by selection) to allow secure individual BPT classification,
averaging the measurements results in strong and weak central sources
landing in the AGN (Seyfert) part of the BPT diagram, and those with
no central source in the SF/AGN composite region. Finally, galaxies
with no extended UV, where we can measure central UV color have ${\rm
  FUV}-{\rm NUV}<0.4$, which is more consistent with a blue continuum
source \citep{agueros} than an old stellar population.

\section{Conclusions}

Early-type galaxies with strong far-UV excess are dominated by SF,
except when they are red in the ${\rm NUV}-r$, in which case the FUV
comes from a compact source, possibly an optically weak AGN. Extended
SF around ETGs has recently been found in some {\it individual} nearby
galaxies, but these have stellar masses one to two orders of magnitude
smaller \citep{thilker10,donovan}. We extended SF in galaxies up to
$M_*\sim 10^{11} M_{\odot}$. However, strong UV excess is rare in yet
more massive ETGs--possible consequence of AGN feedback and a
reflection of an `E-E' dichotomy. In the majority of cases the SF gas
is probably external in origin (minor mergers or IGM accretion),
making these ETGs `rejuvenated'. Most common UV morphology are rings,
which likely result from resonances (from yet undetected bars or other
disturbances). A smaller fraction of UV-excess ETGs may be more
typical green-valley galaxies, where we see the declining original SF
activity. Similar conclusions regarding the heterogeneity of
evolutionary paths of transiting galaxies (but not necessarily ETGs)
were recently reached by \citet{cortese} based on HI content of nearby
($<25$ Mpc) galaxies.

\acknowledgments We thank Stefano Zibetti for help with ADAPTSMOOTH
and Ray Lucas for help with ACS data. Based on observations made with
the NASA/ESA Hubble Space Telescope and supported with NASA grant
HST-GO-11158.03.

\end{document}